\documentclass[showpacs,amssymb,twocolumn,aps]{revtex4}
\usepackage{amssymb}
\usepackage{amsmath}
\usepackage[pdftex]{graphicx}
\usepackage{epstopdf}
\usepackage[utf8]{inputenc}
\usepackage{fancyhdr}
\usepackage{slashed}
\usepackage{turnstile}

\begin{document}

\date{\today}
\title{Hard thermal loops in static background fields}
\author{F. T. Brandt, %\footnote{fbrandt@usp.br}, 
J. Frenkel %\footnote{jfrenkel@fma.if.usp.br}
and J. B. Siqueira }
%\footnote{joao@fma.if.usp.br}}
\affiliation{ Instituto de F\'{\i}sica,
Universidade de S\~ao Paulo,
S\~ao Paulo, SP 05315-970, Brazil}
\begin{abstract}
We discuss the high temperature behavior of retarded thermal loops in static external fields.
We employ an analytic continuation of the  imaginary time formalism and use a spectral 
representation of the thermal amplitudes. We show that, to all orders, the leading contributions of static hard thermal 
loops can be directly obtained by evaluating them at zero external energies and momenta.
\end{abstract}

\pacs{11.10.Wx}

\maketitle

\section{Introduction}

In thermal field theory, in order to deal with the infrared singularities which occur at finite temperature, it is necessary to put thermal masses into the zeroth order of a resummed perturbation theory. To this end, one must first calculate the hard thermal loops, where all the external energies and momenta are much smaller than the temperature $T$. These loops yield gauge-invariant contributions, which are in general non-local functionals of the external fields \cite{Frenkel:1990br,Braaten:1990mz}. However, there are two special cases: the static and the long wave-length limits, when the hard thermal amplitudes become local functions, which are independent of the external energies and momenta \cite{Frenkel:2009pi,Brandt:2009ht}. Nevertheless, these two limits give different functions \cite{Francisco:2013vg}. Of special interest is the evaluation in the above limits of causal thermal self-energy functions, which determine the high-temperature behavior of screening lengths 
and plasma frequencies \cite{Blaizot:2001nr}.

The purpose of this work is to derive a simple method for calculating the leading contributions 
of  retarded thermal loops in static external bosonic fields.
We will show that these contributions can be directly obtained by evaluating the hard thermal loops at zero external energies and momenta. 
This result has been previously derived  in gauge theories at one-loop level
\cite{Frenkel:2009pi} and verified by explicit calculations at two-loops \cite{Brandt:2012mn}.
Here, we present an argument which is valid %, for hard thermal loops, 
to all orders in thermal perturbation theory.

The above result may be more readily understood in the analytically continued imaginary-time formalism 
\cite{kapusta:book89, lebellac:book96, das:book97}
 which is well suited for the study of retarded 
Green's functions \cite{Evans:1991ky}.
This formalism of thermal field theory defines the bosonic Green's functions at integral values of ${k_j}_0/2\pi\, i T$, where ${k_j}_0$ is the energy of the $j$-th  external particle.  After performing the sums over 
the integral values (half-integral for fermions) of ${Q_l}_0/ 2\pi\, i T$, 
where ${Q_l}_0$ is the energy of the $l$-th  internal particle, one arrives at bosonic (fermionic) thermal distribution functions of the form
\begin{equation}\label{eq1}
N(k_0+Q) = \frac{1}{{\rm e}^{(k_0+Q)/T} \mp 1}
\end{equation}
where $k_0$ is some linear combination of external energies and $Q$ is some combination of external and internal momenta.

It is worthwhile to note that if one would now analytically continue the external energies, 
one would get an analytic  behavior when all ${k_j}_o$ and $\vec k_j$ become small, leading to a well defined result in the limit
${k_j}_\mu = 0$. However, the analytic continuation constructed in this manner would yield, after performing the integrations over internal momenta, factors like $\exp{({k_0}/T)}$ which are exponentially increasing for large values of  ${k_0}$.
The proper procedure which avoids the appearance of such factors makes use of the relation
\begin{equation}\label{eq2}
N(k_0+Q) = N(Q)
\end{equation}
which is valid before analytic continuation, since then, ${k_0}/2\pi\, i T$ is an integer.

In this way, the Green's functions will be well behaved when ${k_j}_0$ are analytically continued to complex values, and various 
limits approaching the real axis from different directions may be taken. But this procedure, in contrast to the previous one, introduces
non-analyticities in the thermal loops when ${k_j}_0\rightarrow 0$ and ${\vec k}_j\rightarrow 0$. Nevertheless, the leading 
contributions of hard thermal loops in the static case ${k_j}_0 =  0$, ${\vec k}_j\rightarrow  0$, still agree with those obtained
by setting directly in the loops all external energies and momenta  ${k_j}_\mu  = 0$. This agreement occurs only in the 
static limit which entails that $k_0=0$, since then the condition \eqref{eq2} reduces to an identity. 
Thus, in the static case, analytic continuation preserves the form of the original thermal distribution functions
\eqref{eq1}, which lead to an analytic behavior when all ${\vec k}_j\rightarrow 0$.

We will first exemplify the above argument in the case of a two-loop thermal amplitude. Next, we shall present a more general
approach, based on a spectral representation of the thermal Green's functions \cite{landsman1:1987uw,Evans:1990hy,Taylor:1993ub},  
which allows to verify this argument to all orders.

\section{Thermal self-energy at two-loops}

\begin{figure}[h]
\begin{center}
\includegraphics[scale=0.389775]{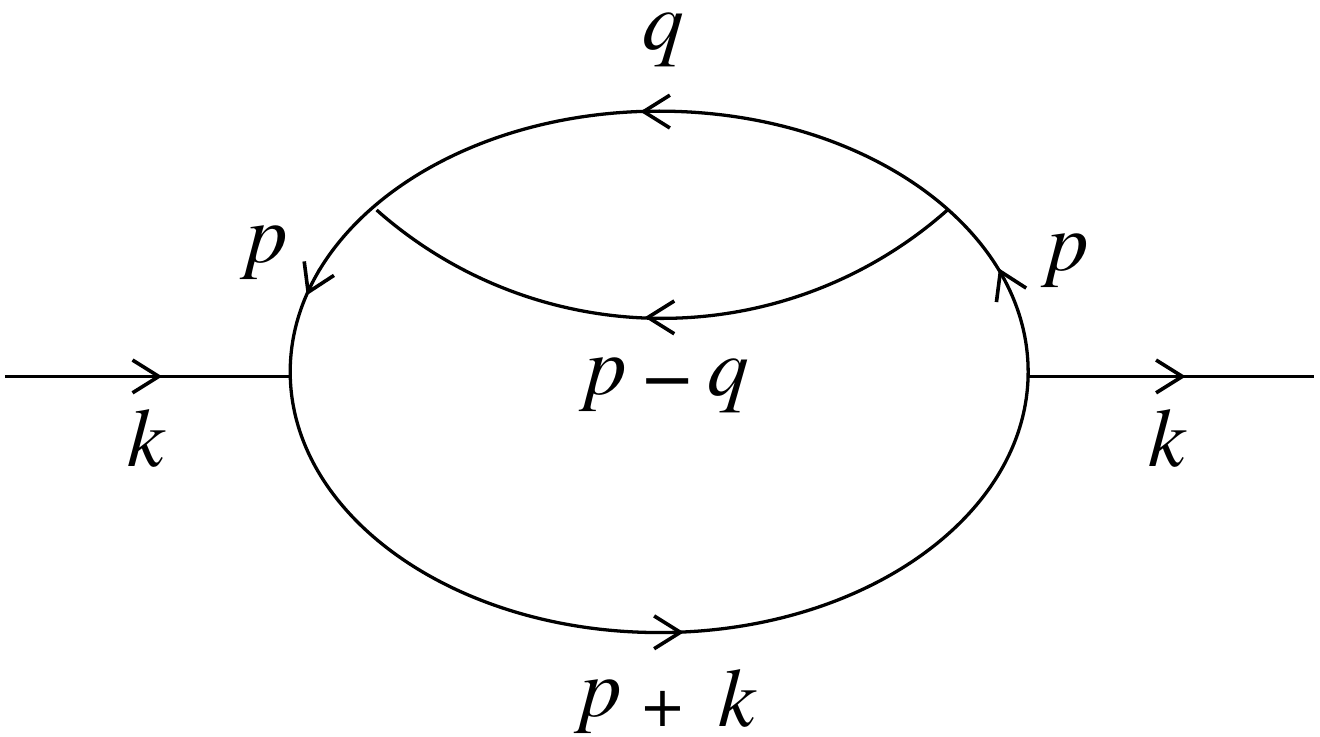}
\caption{A two-loop configuration of the thermal self-energy.}
\label{fig1}
\end{center}
\end{figure}

Let us consider, for example, a two-loop diagram in the scalar  $\lambda \phi^3$ theory, as shown in figure \ref{fig1}.
The result for the one-loop sub-diagram in the renormalizable six-dimensional theory may be written in the 
\hbox{form ($p\equiv |\vec p|$) \cite{Brandt:1991fs}}
\begin{equation}\label{eq3}
\Sigma^{(1)}(p_0,\vec p) = \frac{1}{2}
(p_0^2 - p^2) \Pi_T(p_0,p) + \sum_{n=0}^\infty \frac{P_{2(n+1)}(p_0,p)}{T^{2n}}
\end{equation}
where $P_{2(n+1)}$ is a polynomial of degree $2(n+1)$ in $p_0$ and $p$
(when $n=0$, there appear additional logarithmic terms) and $\Pi_T(p_0,p)$ is given by
\begin{widetext}
\begin{equation}\label{eq4}
\Pi_T(p_0,p) = \frac{\lambda^2}{48\pi} \left\{
\frac{T^2}{p^2}\left[1-\frac{p_0}{2 p}
  \log\left(\frac{p_0+p}{p_0-p}\right)\right]-\frac{1}{4\pi^2} \log \frac{T}{\mu}
\right\},
\end{equation}
\end{widetext}
where $\mu$ is a renormalization mass scale.
The above one-loop result for $\Pi_T(p_0,p)$ has branch points at $p_0=\pm p$. When this term is inserted in the two-loop graph of
Fig. \ref{fig1}, these branch points coincide with the poles at $p_0=\pm p$ which arise from the other terms in this diagram. Therefore, in order
to perform the $p_0$ integration in a well defined manner, it is necessary to regularize $\Pi_T$, which we do using dimensional regularization
in $6+2\epsilon$ dimensions. This leads to the regularized form of  $\Pi_T(p_0,p)$ given by
\begin{widetext}
\begin{equation}\label{eq4a}
\Pi_T(p_0,p,\epsilon) = \frac{\lambda^2}{48\pi} \left\{
\frac{T^2}{p^2}\left[1-\left(1+ {\cal O}(\epsilon)\right)
\frac{p_0}{2}   \int_0^\pi d\theta \frac{\left(\sin\theta\right)^{1+2\epsilon}}{p_0 - p\cos\theta} \right]-\frac{1}{4\pi^2} \log \frac{T}{\mu}\right\},
\end{equation}
which reduces to Eq. \eqref{eq4} in the limit $\epsilon\rightarrow 0$. In Eq. \eqref{eq4a}, $\theta$ denotes the angle between the momenta
$\vec p$ and $\vec q$ shown in Fig. \ref{fig1} and ${\cal O}(\epsilon)$ denote terms which are well behaved at $p_0=\pm p$.
Using this result in the two-loop graph of Fig. \ref{fig1}, and performing the sum over the integral values of $p_0/2 \pi i T$ 
by contour integration, we obtain a leading thermal contribution of the form 
\begin{equation}\label{eq5}
\Sigma^{(2)}_T(k_0,\vec k) = \frac{\lambda^2}{2} \int \frac{d^{5+2\epsilon} p}{(2\pi)^{5+2\epsilon}}\frac{1}{2\pi i} \oint_{C} dp_0 N(p_0) 
\left[
\frac{1}{p_0^2-p^2}\frac{1}{(p_0+k_0)^2-(\vec p+\vec k)^2} \Pi_T(p_0,p,\epsilon) + (p_0\rightarrow -p_0)
\right] 
\end{equation}
\end{widetext}
where the anticlockwise contour $C$,  along the imaginary $p_0$-axis, is closed in the right half $p_0$ plane. 
Evaluating the $p_0$-integral  in terms of the poles inside $C$, and using the fact that the leading contribution from the pole at $p_0=p\cos\theta$
in $\Pi_T(p_0,p,\epsilon)$ vanishes, leads to the result ($\tilde p\equiv |\vec p + \vec k|$)
\begin{widetext}
\begin{eqnarray}\label{eq6}
\Sigma_T^{(2)}(k_0,\vec k) &=& \frac{\lambda^2}{8} \int \frac{d^{5+2\epsilon}p}{(2\pi)^{5+2\epsilon}}\frac{1}{p\tilde p}
\left[
\frac{N(k_0 + \tilde p)\Pi_T(k_0+\tilde p,p) - N(p) \Pi_T(p,p,\epsilon)}{k_0+\tilde p - p} - \right. \nonumber \\  
&& \qquad\qquad\qquad\qquad
\left. \frac{N(k_0+\tilde p)\Pi_T(k_0+\tilde p,p) + N(p) \Pi_T(p,p,\epsilon)}{k_0+\tilde p + p}
+(k_0\rightarrow -k_0)\right].
\end{eqnarray}

At this point, we note that if one would analytically continue \eqref{eq6} to complex values of $k_0$, one would obtain an
expression which is analytic as $k_\mu\rightarrow 0$, with the limit
\begin{equation}\label{eq7}
\Sigma_T^{(2)}(k_0=0,\vec k=0) = \frac{\lambda^2}{4}\int \frac{d^{5+2\epsilon} p }{(2\pi)^{5+2\epsilon}}\frac{1}{p^2} 
\left\{
\left[ N(p_0) \Pi_T(p_0,p,\epsilon) \right]^\prime_{p_0=p} - \frac{N(p) \Pi_T(p,p,\epsilon)}{p}\right\}  ,
\end{equation}
\end{widetext}
where the prime denotes a derivative with respect to $p_0$. 
However, for the reasons mentioned previously, one must first use in \eqref{eq6} the relation
\begin{equation}\label{eq8}
N(k_0+\tilde p) = N(\tilde p)
\end{equation}
and then make the analytic continuation of $k_0$. This procedure will introduce a non-analyticity when $k_\mu\rightarrow 0$. 
To study this, we remark that the leading contribution in $T$ of the $p$-integral in \eqref{eq6} comes from the region where 
$p\sim T$. Thus, in order to find the high-temperature behavior of the hard thermal loop, one may assume that 
$|k_0|$,  $|\vec k| \ll p$. Then, it is easy to see that the leading contribution obtained in the static limit agrees with the result
given in \eqref{eq7} 
\begin{equation}\label{eq9}
\Sigma_T^{(2)}(k_0=0,\vec k\rightarrow 0) = \Sigma_T^{(2)}(k_0=0,\vec k =  0). 
\end{equation}
On the other hand, the leading contributions in the long wave-length limit would lead to a different result, namely
\begin{widetext}
\begin{equation}\label{eq10}
\Sigma_T^{(2)}(k_0 \rightarrow 0,\vec k = 0) = \frac{\lambda^2}{4} \int \frac{d^{5+2\epsilon} p  }{(2\pi)^{5+2\epsilon}}\frac{1}{p^2} 
N(p)\left[
\left.\Pi_T^\prime(p_0,p,\epsilon)\right|_{p_0=p} - \frac{\Pi_T(p,p,\epsilon)}{p}\right]. 
\end{equation}
\end{widetext}
%Similar relations also hold for the other two-loop configuration of the self-energy function. 

Let us now evaluate the leading static thermal contribution which arises from \eqref{eq9}. 
In this case, there appear individual terms proportional to $1/\epsilon$
which exhibit a collinear singularity in the region where $p_0^2=p^2$, 
with $\vec p$ and $\vec q$ being nearly parallel (see Fig. \ref{fig1}).
However, such collinear singularities turn out to cancel 
so that in the present case, it is not necessary to resort to the KLN mechanism at finite
temperature \cite{Grandou:1991qr}.
Using Eqs. \eqref{eq4a} and \eqref{eq7}, together with the relations
\begin{equation}\label{eq11}
\int_0^\infty dp N(p) p = -\frac{1}{2}\int_0^\infty dp N^\prime(p) p^2 = \frac{\pi^2 T^2}{6}
\end{equation}
one then gets a leading $T^2 \log T$ contribution of the form
\begin{equation}\label{eq12}
\Sigma_T^{(2)}(k_0=0,\vec k\rightarrow 0) =
\frac{\lambda^4}{144}\frac{\pi}{4}\frac{1}{(2\pi)^5} T^2 \log \frac{T}{\mu}.
\end{equation}

\section{The self-energy to all orders}

In order to verify to all orders that the leading static contributions agree with
the result obtained by setting directly, in the self-energy, the external energy-momentum equal to zero,
it is convenient to use a spectral representation of the
analytically continued bosonic self-energy 
(see, for example, chapter 3 in \cite{landsman1:1987uw} and references therein).
%This is obtained by using an analytic continuation of the exact propagator $\Delta(z,\vec k)$, which is unique provided one requires
%it to be analytic outside the real axis and to vanish when $|z|\rightarrow\infty$. This analytic propagator has no complex zeros off
%the real axis, which implies that the inverse propagator may also exist as an analytic function. Using the relation between the inverse
%propagator and the bosonic self-energy $\Sigma(z,\vec k)$, and separating off the contribution at infinite
%frequency $\Sigma(\infty,\vec k)$, one arrives at the relation
One then obtains for the analytic retarded self-energy the spectral form
\begin{equation}\label{eq14}
\Sigma(k_0,\vec k) =  \Sigma(\infty,\vec k) + \int_{-\infty}^{\infty} \frac{d\kappa_0}{2\pi}\frac{\sigma(\kappa_0,\vec k)}{k_0-\kappa_0+i\epsilon},
\end{equation}
where $k_0$ is a real energy and, for simplicity, only the energy-momentum dependence has been written explicitly.
Here, the spectral density
$\sigma(\kappa_0,\vec k)=-\sigma(-\kappa_0,\vec k)$  is related to the discontinuity of $\Sigma(\kappa_0,\vec k)$
across the real axis 
%Specifically, we will consider the retarded self-energy function, which is obtained by setting $z=k_0+i\epsilon$, where $k_0$ is a positive
%frequency.
%Then, 
and the second term %in Eq. \eqref{eq14} is analytic outside the real axis and 
approaches zero when $|k_0|\rightarrow\infty$. 
These features can be easily seen at one-loop in the scalar $\lambda \phi^3_6$ model, where $\Sigma(\infty,\vec k) =0$ and
\begin{widetext}
\begin{equation}\label{eq15}
\sigma(\kappa_0,\vec k)=-\frac{\lambda^2}{4\pi} \frac{1}{(2\pi)^6}  \int {d^6 p} \int {d^6 p^\prime}
\rho_0(p) \rho_0(p^\prime) \delta^6(p+p^\prime+\kappa)\left[1+N(p_0)+N(p_0^\prime) \right].
\end{equation}
\end{widetext}
Here, $\kappa_\mu=(\kappa_0,\vec k)$ and the free spectral density $\rho_0(p)$ is given by
\begin{equation}\label{eq16}
\rho_0(p) = 2 \pi \epsilon(p_0) \delta(p_0^2-p^2-m^2),
\end{equation}
where $m$ is the ordinary mass of the scalar particles. One can check that, by
substituting \eqref{eq15} into \eqref{eq14} and performing the $\kappa_0$ integration,  one gets the one-loop thermal contribution
\begin{widetext}
\begin{equation}\label{eq17}
\Sigma_T^{(1)}(k_0,\vec k) = \frac{\lambda^2}{8}\int \frac{d^5p }{(2\pi)^5}\frac{1}{p\tilde p}
\left[
\frac{N(\tilde p) - N(p)}{k_0+\tilde p-p} - \frac{N(\tilde p) + N(p)}{k_0+\tilde p+p} + (k_0\rightarrow -k_0)
\right] ,
\end{equation}
\end{widetext}
where $k_0 \rightarrow k_0 + i\epsilon$ is to be understood and we have neglected $m$ with respect to $p$.
This agrees with the result obtained from \eqref{eq6} by setting $\Pi_T=1$ and employing the relation \eqref{eq8}.
Evaluating \eqref{eq17} in the static and long wavelength limits, leads to distinct terms of order $T^2$.

Using the spectral representation \eqref{eq14}, where we take for definiteness $\Sigma(\infty,\vec k)=0$, we now consider the leading
thermal contributions which arise in the long wave-length and static limits of the retarded self-energy
\begin{equation}\label{eq18}
\Sigma_T(k_0\rightarrow 0,\vec k=0)=\frac{1}{2\pi} \lim_{k_0\rightarrow 0}\int_{-\infty}^\infty \frac{d \kappa_0}{k_0-\kappa_0+i\epsilon}\sigma_T(\kappa_0,0),
\end{equation}
\begin{equation}\label{eq19}
\Sigma_T(k_0 = 0,\vec k\rightarrow 0)=-\frac{1}{2\pi} \lim_{\vec k \rightarrow 0}
\int_{-\infty}^\infty \frac{d \kappa_0}{\kappa_0-i\epsilon}\sigma_T(\kappa_0,\vec k).
\end{equation}
Next, let us compare these contributions with the result obtained by setting directly $k_0=0$, $\vec k=0$ in the retarded thermal self-energy function
\begin{equation}\label{eq20}
\Sigma_T(k_0 = 0,\vec k = 0)=-\frac{1}{2\pi} \int_{-\infty}^\infty \frac{d \kappa_0}{\kappa_0-i\epsilon}\sigma_T(\kappa_0,0).
\end{equation}
Since the integrand in $\Sigma_T(k_0,\vec k = 0)$ is not a uniformly continous function of $k_0$,
% analytic  on the real $k_0$ axis, 
we  cannot take the limit $k_0\rightarrow 0$ inside the integral \eqref{eq18}. Thus, we
infer that the leading thermal contribution \eqref{eq18} got in the long wave-length
limit, would generally differ from the result given in \eqref{eq20}. 

On the other hand, we will argue that in the static case, the limit $\vec k\rightarrow 0$ can be taken inside the integral
\eqref{eq19}. To this end, using the fact that $\sigma_T(\kappa_0,\vec k)$ is an odd function of $\kappa_0$, it is convenient to write 
\eqref{eq19} in the alternative form (where ${\cal P}$ denotes the principal value) 
\begin{equation}\label{eq21}
\Sigma_T(k_0 = 0,\vec k\rightarrow 0)=-\frac{1}{\pi} \lim_{\vec k \rightarrow 0} {\cal P}
\int_{0}^\infty \frac{d \kappa_0}{\kappa_0}\sigma_T(\kappa_0,\vec k)
\end{equation}
Consider now the integral
\begin{equation}\label{eq22}
I(\vec k)={\cal P} \int_{0}^\infty \frac{d \kappa_0}{\kappa_0}\sigma_T(\kappa_0,\vec k).
\end{equation}
It is well known that if $I(\vec k)$ converges uniformly, then the limit $\vec k\rightarrow 0$ can be taken inside the integral.

%On the other hand, 
%based on lower order calculations, we will assume that $\sigma_T(\kappa_0,\vec k)$ is a well behaved function when 
%${\vec k \rightarrow 0}$, such that this limit may be taken inside the integral \eqref{eq19}. 

Such a convergence may be shown by considering the physical meaning of $\sigma_T(\kappa_0,\vec k)$, which gives the imaginary
part of the retarded self-energy. 
It yields the rates of processes occurring in a thermal plasma, such  as particle 
creation/annihilation or scattering, in the presence of an external field \cite{lebellac:book96}.
At high temperatures, such that $|\vec k | \ll T$, the leading contributions to these rates %are expected to 
have  
a smooth behavior when $\vec k\rightarrow 0$, in which case $\sigma_T(\kappa_0,\vec k)$ would be a well behaved function in this limit.
%understood by considering
%the behavior of the leading static thermal contributions in $\Sigma_T(k_0=0,\vec k \rightarrow 0)$. In this case, a smooth 
%$\vec k\rightarrow  0$ limit of the static self-energy should be expected    to
%exist in a theory with a well defined Debye screening length.
Furthermore, in consequence of unitarity (conservation of probability) such rates should decrease at large values of the energy $\kappa_0$.
Assuming, for example, that
$\sigma_T(\kappa_0,\vec k)$ behaves for large $\kappa_0$ 
like $\kappa_0^{2} \exp{(-C \kappa_0/T)}$, where $C$ is a positive constant
(which is consistent with \eqref{eq15}), one gets
\begin{equation}\label{eq23}
\left |I(\vec k) -{\cal P} \int_{0}^E \frac{d \kappa_0}{\kappa_0}\sigma_T(\kappa_0,\vec k)  \right| 
\propto \frac{E}{T}\exp{-\frac{CE}{T}} < \epsilon
\end{equation}
for every $\epsilon$, provided $E/T$ is sufficiently large. 
A similar condition is obtained also for more general forms of the spectral density at large $\kappa_0$, which lead to a $T^2$ behaviour of the self-energy at
high temperatures.
Thus, $I(\vec k)$ will be uniformly convergent so that the
limit $\vec k\rightarrow 0$ can be taken inside  \eqref{eq21} and, therefore, inside the integral $\eqref{eq19}$.

Consequently, it follows that the leading static contribution \eqref{eq19}  will
agree to all orders with the result \eqref{eq20}, got by 
calculating $\Sigma_T$ at vanishing energy and momentum. 
This behavior is in agreement with the two-loop results given in Eqs. \eqref{eq9} and \eqref{eq12}.

\section{generalization to $n$-point functions}
According to the simple arguments
given following Eqs. \eqref{eq1} and \eqref{eq2}, the above result should hold as well in the case of higher point functions.
We will now derive this property for the $n$-point Green's functions, 
using a treatment which generalizes the previous method. %to all higher point functions. 

The spectral representation of the retarded $n$-point functions calculated in the imaginary time formalism may be written 
in the form \cite{Evans:1990hy,Evans:1991ky}
\begin{widetext}
\begin{eqnarray}\label{eq24}
\Gamma^{(n)}\left(\left\{{k_j}_0,\vec k_j \right\}\right) &=&
\left(\frac{-1}{2\pi}\right)^{n-1}\int_{-\infty}^{\infty} \ d{\kappa_1}_0 \dots d{\kappa_n}_0 \delta({\kappa_1}_0 + \dots + 
{\kappa_n}_0) \left[ \rho_{12\dots n}\left(\left\{\kappa_j\right\},T\right) \right. \nonumber \\ 
&\times & 
\frac{i}{{k_2}_0+i\epsilon_2 + {k_3}_0+i\epsilon_3 + \dots + {k_n}_0 + i\epsilon_n 
- \left({\kappa_2}_0 + {\kappa_3}_0 + \dots + {\kappa_n}_0 \right)} 
\nonumber \\ 
&\times & 
\frac{i}{{k_3}_0+i\epsilon_3 + \dots + {k_n}_0 + i\epsilon_n - \left({\kappa_3}_0 + \dots + {\kappa_n}_0 \right)}
\times \dots \times 
\frac{i}{{k_n}_0+i\epsilon_n - {\kappa_n}_0} \nonumber \\
&+& \left. \mbox{all permutations of } (1,2,3,\dots , n) \right]
\end{eqnarray}
\end{widetext}
where ${k_j}_0$ are real energy variables and
${\kappa_j}_\mu = ({\kappa_j}_o,{\vec k}_j)$. Setting 
in \eqref{eq24}, for example, $\epsilon_l$  positive and all other epsilons negative such that
$\sum \epsilon_j = 0$, defines the analytic $l$-th retarded function. The spectral densities are the difference of two thermal
Wightman functions which, in the case of pure bosonic fields, are given by
\begin{widetext}
\begin{eqnarray}\label{eq25}
\rho_{12\dots n}\left(\left\{\kappa_j\right\},T\right) & = &
{\rm Tr} \left\{{\rm e}^{-H/T} \phi_1(\kappa_1) \phi_2(\kappa_2) \dots \phi_n(\kappa_n)
\right\}/ {\rm Tr} \left\{{\rm e}^{-H/T}\right\} 
\nonumber \\
&&\!\!\!\!\!\!\!\!\!\!\!\!\!\!\!\!\!\!\!
-(-1) ^n {\rm Tr} \left\{{\rm e}^{-H/T} \phi_n(\kappa_n) \dots \phi_2(\kappa_2) \phi_1(\kappa_1)\right\}/ {\rm Tr} \left\{{\rm e}^{-H/T}\right\} .
\end{eqnarray}
\end{widetext}
We assume that to leading order at high temperature, when all $|\vec k_j|\ll T$, 
these spectral densities (which may also depend on the particles masses, etc) are well behaved in the limit
$\vec k_j \rightarrow 0$.

One may now consider the leading thermal contributions which arise from \eqref{eq24} in the static limit 
(when all ${k_j}_0=0$)  which is well defined due to 
the analyticity properties of $\Gamma^{(n)}$. In this case, to leading order in $T$, one may next take the limits 
$\vec k_j\rightarrow 0$ and proceed similarly  to the previous analysis.  We then find a result which agrees with
that obtained by setting in \eqref{eq24} all external energies and momenta equal to zero:
\begin{widetext}
\begin{eqnarray}\label{eq26}
\Gamma^{(n)}_T\left(\left\{{k_j}_0=0,\vec k_j \rightarrow 0\right\}\right) &=&
\left(\frac{i}{2\pi}\right)^{n-1}\int_{-\infty}^{\infty} \ d{\kappa_1}_0 \dots  d{\kappa_n}_0 
\delta({\kappa_1}_0 + \dots + {\kappa_n}_0)
\left[ \rho_{12\dots n}\left(\left\{{\kappa_j}_0,\vec k_j=0\right\},T\right) \right. \nonumber \\ &\times & 
\frac{1}{{\kappa_2}_0 + {\kappa_3}_0 + \dots+ {\kappa_n}_0 - i(\epsilon_2+\epsilon_3+\dots+\epsilon_n)}
\nonumber \\ &\times & 
\frac{1}{{\kappa_3}_0+\dots+{\kappa_n}_0-i(\epsilon_3+\dots+\epsilon_n)} \times \dots \times
\frac{1}{{\kappa_n}_0-i\epsilon_n} 
\nonumber \\ & + & \left.
\mbox{all permutations of } (1,2,3,\dots,n)\right] .
\end{eqnarray}
\end{widetext}
We have explicitly verified this relation at two-loops order, by calculating the leading static thermal contributions of the three-point functions in gauge
theories. Thus, we conclude that to all orders in the static limit, the leading thermal contributions of retarded bosonic Green's functions may be directly obtained by
evaluating them at zero external energies and momenta.

\acknowledgments
We would like to thank FAPESP and CNPq (Brazil) for a grant.
%J. F. is indebted to Prof. J. C. Taylor for a helpful correspondence.

%\bibliography{all_new}

\end{document}